\documentclass[aps,pra,twocolumn,showpacs,floatfix,footinbib,groupedaddress,superscriptaddress]{revtex4}
\usepackage{amsmath}
\usepackage{amssymb,graphics,graphicx,times,bm,bbm,multirow}
\usepackage{footnote}
\usepackage{appendix}

\begin{document}

\title{Numerical Evidence for Robustness of Environment-Assisted Quantum Transport}

\author{A. Shabani}
\affiliation{Department of Chemistry, Princeton University,
Princeton, New Jersey 08544}
\author{M. Mohseni}
\affiliation{Center for Excitonics, Research Laboratory of
Electronics, Massachusetts Institute of Technology, Cambridge, MA
02139}
\author{H. Rabitz}
\affiliation{Department of Chemistry, Princeton University,
Princeton, New Jersey 08544}
\author{S. Lloyd}
\affiliation{Department of Mechanical Engineering, Massachusetts
Institute of Technology, Cambridge, MA 02139}

\begin{abstract}
Recent theoretical studies show that decoherence process can enhance transport efficiency in quantum systems.
This effect is known as environment-assisted quantum transport (ENAQT). The role of ENAQT in optimal quantum transport
is well investigated, however, it is less known how robust ENAQT is with respect to variations in the system or its environment characteristic.
Toward answering this question, we simulated excitonic energy transfer in Fenna-Matthews-Olson (FMO) photosynthetic complex.
We found that ENAQT is robust with respect to many relevant parameters of environmental interactions and
Frenkel-exciton Hamiltonian including reorganization energy, bath frequency cutoff, temperature, and initial excitations, dissipation rate,
trapping rate, disorders, and dipole moments orientations.
Our study suggests that the ENAQT phenomenon can be exploited in robust design of highly efficient quantum transport systems.
\end{abstract}

\maketitle

\section{Introduction}

Quantum or coherent transport are common phenomena in many natural or artificial systems \cite{MayBook,DattaBook,QEBbook}.
In a quantum system, transfer of matter or energy is strongly influenced by structural disorder or environmental noise. 
Recently, it was discovered, in the context of energy transfer in photosynthetic complexes, that environmental interactions can have
a constructive role in excitonic transport, called ENAQT effect \cite{mohseni-fmo,Rebentrost08-1,Rebentrost08-2,mohseni13,Plenio08-1,Plenio09,Castro08,Chin13,Olaya13,QEBbook}. Basically, a decoherence process can facilitate excitons displacement between molecules by breaking wavefunction localization \cite{Rebentrost08-1,mohseni13}, energy level broadening \cite{Plenio08-1,Plenio09}, or quantum jumps \cite{mohseni-fmo,Castro08}. Based on ENAQT theory, optimal quantum transport is achieved by an optimal interplay of the system coherence and decoherence dynamics, influenced by the environment structure \cite{Chin13,Olaya13}. The convergence of system-environment energy scales is an underlying principle for such an optimal transport process \cite{mohseni-fmo}.

In this paper, we study the role of ENAQT in the robustness of quantum transport, a feature complement to optimality.
To this end, as a model system, we consider excitonic energy transfer in FMO complex and investigate the efficiency of transport in the presence of large variations in internal and external conditions.
The FMO complex is a trimer consisting of three identical monomers each formed from seven bacteriochlorophylls (BChl) embedded in a scaffold protein. An FMO unit acts as an energy transfer channel in green sulphur bacteria
guiding excitons from the light-harvesting antenna complex to the reaction center.
Recent electronic spectroscopy experiments provide
evidence that long-lived quantum dynamical coherence can exist in
FMO complex \cite{Engel07,Lee07,panit10,Engel12,Engel12-2}.
Such experimental observation suggests that modeling excitonic dynamics
requires describing FMO as an open quantum system to include both the internal
coherence and the decoherence induced by the protein scaffold environment.  

For our modeling, we employed the time-convolutional master equation (TC2) that we derived and analyzed in Ref.\cite{Shabani11}.
In our study, we consider the environmental parameters
including reorganization energy, bath cutoff frequency, temperature, and trapping. The role of antenna complex is studied by varying the initial
excitonic states and its impact on energy transfer efficiency (ETE). For the internal parameters, we study the robustness of ETE in presence of disorder in the
FMO internal structure parameterized by site energies,
inter-chlorophyl distances, and dipole moment orientations.
We observe that ENAQT enhances robustness of energy transfer while is a universal phenomenon in the sense that the environment can assist transport even at non-optimal regime of parameters.

We should mention that the results presented in this manuscript are part of an extensive study that we had reported its other aspects in Refs.\cite{Shabani11,mohseni13,mohseni13-2}. A new derivation of the TC2 master equation along with assessment of its reliability for calculating energy transfer efficiency beyond perturbative and Markovian limits was presented in \cite{Shabani11}. Ref. \cite{mohseni13} presents an underlying principle, the convergence of system and environment energy-scales to describe optimal quantum transport. We discussed the role of geometrical character of an excitonic system in energy transfer in Ref. \cite{mohseni13-2}. The current study presents numerical evaluation of robustness for ENAQT phenomenon while our previous papers \cite{Shabani11,mohseni13,mohseni13-2} were addressing the role of ENAQT is enhancement of energy transfer.

\section{Theoretical model}

\begin{figure}[tp]
\includegraphics[width=9cm,height=6cm]{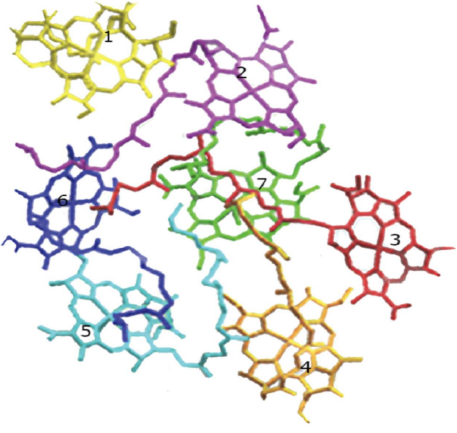}
\caption{The disordered structure of the Fenna-Matthews-Olson (FMO)
complex. Seven BChls are embedded in a protein scafold that is modeled as a bath
of harmonic oscillators. Initial exciton is usually formed on BChls 1 and 6 that are in
proximity of light-harvesting antenna complex, and is transferred to the reaction center which is in the
proximity of BChls 3 and 4.} \label{Figfmo}
\end{figure}

The FMO electronic states at low energy limit is modeled as a $7$-level quantum system with Hamiltonian

\begin{equation}
H_{S} =\sum_{j,k}\epsilon _{j}|j\rangle \langle j|+J_{jk}|j\rangle
\langle k|,
\end{equation}
where $|j\rangle $ denotes an excitation state in a
chromophore spatially located at site $j$. The diagonal site
energies are denoted by $\epsilon _{j}$s. The
strengths of dipole-dipole interactions between chromophores in
different sites are represented by $J_{jk}$. We model the protein scaffold
as a bath of harmonic oscillators with Gaussian fluctuations \cite{McKenzie}. Each BChl
is coupled to a separate bath with interaction Hamiltonian $H_{SB}=S_jB_j$ where $S_j=|j\rangle\langle j|$, and $B_j$ is the $j$'th bath operator. We assume a Drude-Lorentzian correlation function, given at temperature $T$ as:
\begin{eqnarray}
C_j(t)&=&\langle \tilde{B}_j(t)\tilde{B}_j(0)\rangle\notag\\
&=&\frac{1}{\pi}\int_0^\infty
d\omega J(\omega)\frac{\exp(-i\omega t)}{1-\exp(-\hbar\omega/k_BT)},
\end{eqnarray}
where the spectral function has the form
$J(\omega)=2\lambda\omega/(\omega^2+\gamma^2)$. For FMO,
we considered the reorganization energy value of
$\lambda=35$ $cm^{-1}$ and bath cutoff frequency $\gamma=50$ $cm^{-1}$, all the same for different BChls.

In order to model the FMO electronic degrees of freedom as an open quantum system,
we employ the time-nonlocal master equation TC2.
\begin{eqnarray}
&&\frac{\partial}{\partial t}\rho(t)=\mathcal{L}_S\rho(t)+\mathcal{L}_{e-h}\rho(t) \label{TNME}\\
&&-\sum_j[S_{j},\frac{1}{\hbar^2}\int_0^t C_j(t-t')
e^{\mathcal{L}_S(t-t')} S_{j}\rho(t') dt'-h.c.]\notag
\end{eqnarray}
where the $\mathcal{L}_{S}=-i[H_S,.]$. Here, we include the effect of exciton loss and reaction center (RC) trapping by the superoperator
$\mathcal{L}_{e-h}=-\sum_j r_{loss}^j\{|j\rangle\langle j|,.\}-r_{trap}\{|trap\rangle\langle trap|,.\}$. In the first term, the loss rate is $r_{loss}^j=(1ns)^{-1}$ while
the second term $r_{trap}\{|trap\rangle\langle trap|,.\}$ represents the exciton capturing process by the reaction center.
We consider BChl3 as the trapping site with the trapping rate of $r_{trap}=(0.5ps)^{-1}$.

We quantify ETE as the total portion of a traveling exciton successfully captured by the reaction center:
\begin{eqnarray}
\eta=2r_{trap}\int_0^\infty \langle trap|\rho(t) |trap\rangle dt
\label{ETE}
\end{eqnarray}
The above performance function had been used in our previous studies
\cite{mohseni-fmo,Shabani11,mohseni13,mohseni13-2}. In Ref.\cite{Shabani11}, we examined that TC2 equation provides reliable estimation of ETE for the range of parameters considered in this study.

\section{Robustness with respect to environmental
parameters}

We examine the
degree of optimality and robustness of the energy transfer by
employing the the Euclidean norm of the gradient and Hessian matrix
of the ETE function. The Euclidean norm of the ETE gradient at any
parameters values $p_1$ and $p_2$, $||\nabla{\eta(p_1,p_2)}||_2$,
quantifies the degree of optimality. The gradient measure reveals
the degree of local optimality in a surface manifold. Careful
inspections of the room temperature plots for the ETE function
versus various pairs of relevant parameters show a convex or concave
manifold, thus gradient as a measure of local optimality suffices to
measure global optimality. To examine the robustness, we compute the
Hessian matrix norm $||H(\eta(p1,p2))||
_2=||[\partial^2\eta/\partial p_i\partial p_j]||_2$ ($i,j=1,2$) as
the total  measure of local curvature of the manifold. A smaller
value of this norm corresponds to a flatter surface, thus a more
robust process. We use a five-point stencil method to compute
derivatives numerically.

\begin{figure}[tp]
\includegraphics[width=9cm,height=6cm]{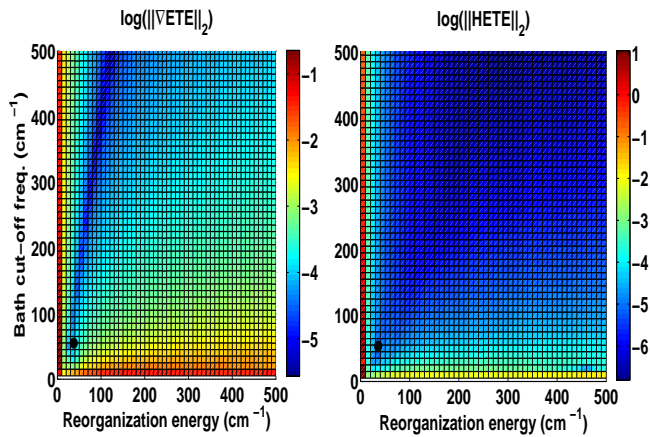}
\caption{Left: The degree of ETE optimality is quantified at different
values of $\lambda$ and $\gamma$ by the gradient matrix norm of the
ETE function. The dark blue points represent near optimal values.
Right: The degree of ETE robustness is quantified by the Euclidian
norm of Hessian of the ETE. The dark blue points represent near
robust points. The estimated FMO environmental values of $\lambda=35
cm^{-1}$ and $\gamma=50$ $cm^{-1}$, marked by black dots, are
located on the corner of both robust and optimal region.}
\label{Figopt-sus-lamgam}
\end{figure}

A quantitative study of the degree of optimality and robustness
of the energy transfer as functions of system-bath coupling strength
and bath memory is illustrated in Fig \ref{Figopt-sus-lamgam}. The
optimality is defined as Euclidean norm of the ETE function gradient
$||\nabla{\eta(\lambda,\gamma)}||_2$ to locate the local maxima in
the ETE landscape in Fig. \ref{Figopt-sus-lamgam}. The robustness is
defined by $||H(\eta(\lambda,\gamma))|| _2$ to measure local
curvature of the manifold. Note that the ETE gradient and Hessian
matrix norms are indicated in a logarithmic scale, thus the global
optimal point with zero derivative can not be explicitly highlighted
in this representation. The experimentally estimated values for FMO
are illustrated as black dots in each graph clearly located in an
optimal and robust region. One remarkable feature is the fact that
environmental parameters of FMO have almost the minimal
reorganization energy and bath cutoff frequency among all the
regions with simultaneous optimal and robust properties. One hypothetical explanation could be the overall tendency in nature to minimize the
amount of required work, that is, facilitating an optimal and robust
environmental platform for the FMO energy transport by preserving a
rather small size scaffold protein that is weakly coupled. However, one may ask
why nature has not evolved toward an even more compact complex.
On reason could be that pigments at closer distances can exchange electrons
in addition to excitons that would reduce the excited state lifetime, the so-called concentration quenching. However, using the modeling considered in this paper we are not able to examine such hypothesis.

\begin{figure}[tp]
\includegraphics[width=9cm,height=6cm]{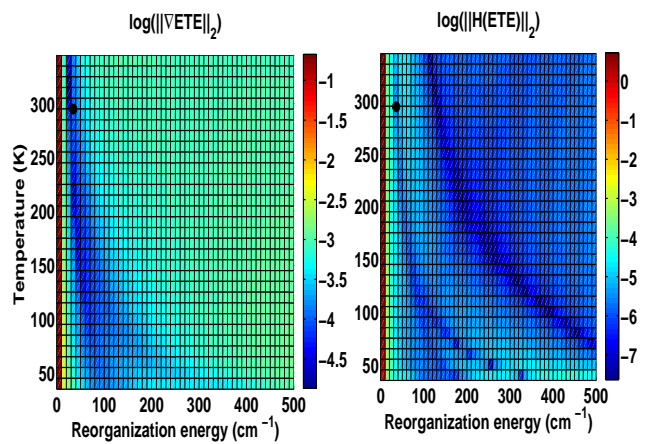}
\caption{ Left: the degree of ETE optimality is quantified at different values
of $\lambda$ and $T$ by the norm of the ETE gradient. The dark blue
points represent near optimal values. Right: the degree of ETE
robustness is quantified by the Euclidian norm of the ETE Hessian.
The dark blue points represent near robust points. We note that
within the range of possible FMO operating temperatures (e.,g
between $T=280^\circ K$ to $T=350 ^\circ K$) simultaneous optimal
and robust energy transport can only be achieved for $\lambda$
values around $30$ to $35$ $cm^{-1}$, that is equivalent to the
estimated reorganization energy for the FMO complex that is marked
by a black dot at room temperature $T=298^\circ K$.}
\label{Figopt-sus-lamT}
\end{figure}

The gradient and Hessian norms as functions of reorganization energy and
temperature are illustrated in Fig. \ref{Figopt-sus-lamT}. At the
relevant FMO operating temperatures, optimum and robust energy
transport can occur simultaneously only within a small regions of
$\lambda$ between $30$ to $35$ $cm^{-1}$, which coincide with the
estimated values of reorganization energy for the FMO. We note that
there are certain regions of higher robustness at higher
reorganization energy that are in principle available, but these
regions imply a significantly lower operating temperature for the
FMO operation and they have suboptimal ETE in comparison with actual
FMO environmental parameters at the room temperature.
The robustness with respect to environmental parameters had been previously reported Refs. \cite{Shabani11,mohseni13,Wu} but with no direct quantification of robustness and mere graphical observation.

\section{Energy transport sensitivity on the initial excitations}

\begin{figure*}[tp]
\includegraphics[width=18cm,height=6cm]{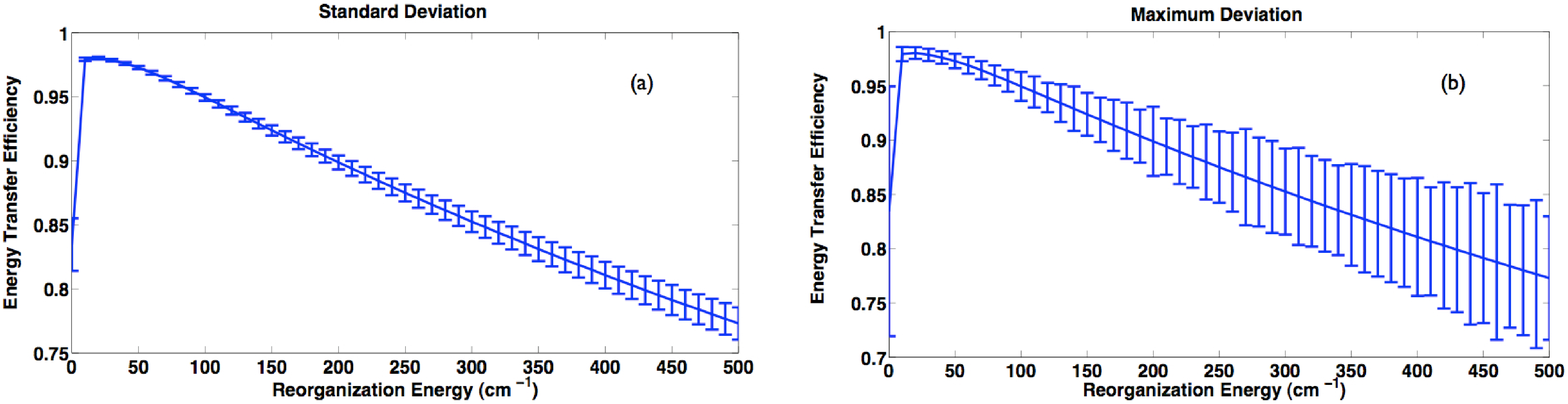}
\caption{The degree of sensitivity of ETE for $10^4$ uniformly
sampled pure and mixed initial exciton density matrices for
different values of reorganization energy: (a) The error bars
indicate the ETE standard deviation. At the FMO value of $\lambda=35
cm^{-1}$ the standard deviation of ETE has a negligible value of
about $0.1\%$. (b) A worst-case scenario of FMO energy transport
sensitivity to initial exciton states. Here, the error bars indicate
the maximum and minimum values of ETE achieved over the sample of
$10^4$ randomly chosen initial states. This plots clearly shows how ENAQT
significantly reduces the dependence of ETE on the initial state of dynamics.}
\label{Figinitial_state_ave}
\end{figure*}

The exciton migration pathways and time-scales have been studied in
detail for a variety of light-harvesting complexes using various
perturbative techniques including F\"{o}rster models for studying
LHI and II of purple bacteria \cite{Ritz} and Lindblad models for
simulating the dynamics of the FMO protein of green sulphur bacteria
\cite{mohseni-fmo}.
%and PSI of higher plants \cite{X}.
Nevertheless, the role of initial conditions in the overall energy
transfer efficiency of photosynthetic complexes is to a large extent
unknown. It was recently shown that the initial quantum coherence
could influence the energy transfer efficiency in LHI of purple
bacteria assuming no interaction with the phonon bath
\cite{Castro08}. The dependency on initial localized excitation at
BChls 1 and 6 were also examined for the FMO complex using Lindblad,
Haken-Strobl, and HEOM models
\cite{mohseni-fmo,Rebentrost08-2,AkiPNAS,Cui}. However, the sensitivity
of ETE with respect to generic initial pure and mixed states taken
from a large ensemble in the single-excitation manifold has not
previously been explored.

\begin{figure}[tp]
\includegraphics[width=9cm,height=6cm]{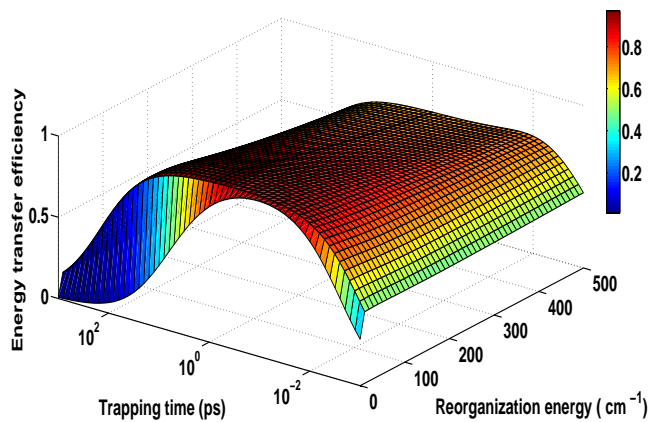}
\caption{The ETE manifold on the parameter space of reorganization
energy and trapping time-scale. It can be seen that the FMO complex
reaches its maximum functionality at trapping rates of about $0.5-5$
$ps^{-1}$. The tunnel-shape ETE landscape can be understood by
noting that at low trapping rates the transport efficiency is
diminished by the recombination process. At high trapping rates the
exciton transfer is suppressed via quantum Zeno effect as the
trapping process corresponds to very strong and continuous
measurement of the system.} \label{Figefftrapcorrect}
\end{figure}

Here, we first examine the average sensitivity of ETE with respect
to randomly chosen initial states for various reorganization
energies. To this end, for each value of reorganization energy, we
sample over $10^4$ (pure or mixed) density matrices from a uniform
distribution in the space of all $7\times 7$ trace one positive
matrices. In Fig.~\ref{Figinitial_state_ave} (a) the average values
of ETE is plotted with an error bar representing the variance of ETE
in our random sampling. Note that at the optimal ETEs, corresponding
to the value of reorganization energy of the FMO complex, the
dependency of the variances on initial states is very small -- less
than $0.1\%$. However, the ETE fluctuations can grow by an order of
magnitude for larger or smaller values of $\lambda$. We also
investigate the best and worst possible random initial single
excitonic states in the Hilbert space of FMO. In
Fig.~\ref{Figinitial_state_ave} (b), we illustrate this extreme
possible deviations by error bars on the average ETE at any given
value of $\lambda$. Note that ETE is very robust, varying about
$1\%$ with respect to different initial excitations at the optimal
area of ETE landscape. However, this robustness diminishes
substantially at the regimes of large reorganization energy. Next,
we study the ETE landscape as a function of trapping and dissipation
rates.

\section{Temporal and geometrical effects of the trapping mechanism}

\begin{figure}[tp]
\includegraphics[width=9cm,height=6cm]{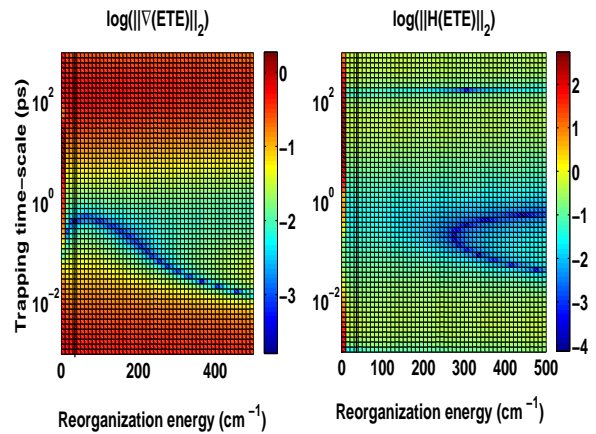}
\caption{This figure is a complement to Fig. \ref{Figefftrapcorrect}
(Left). The degree of ETE optimality is quantified at different
values of $\lambda$ and $r_{trap}^{-1}$ by the norm of the ETE
gradient. The dark blue points represent near optimal values.
(Right) The degree of ETE robustness is quantified by the Euclidian
norm of the ETE Hessian. The blue points represent near robust
points. The FMO achieves its maximum efficiency at
$r_{trap}^{-1}=0.5 ps$. Note that for larger reorganization energies
the trapping rate has be increased to achieve optimal ETE. However,
this competition does not exist at small and intermediate
system-bath coupling strength, where $\lambda$ is on the order of
off-diagonal elements of the FMO free Hamiltonian leading to
environment-assisted energy transport. This suggests that a general
convergence of time-scale might be required to obtain global
efficient and robust transport \cite{mohseni13}.} \label{Figopt-sus-lamtrap}
\end{figure}

Basic structural information on the FMO-RC complex has been obtained
via linear dichroism spectra and electron microscopy
\cite{Remigy99}. These studies indicate that the symmetry axis of
the trimer is normal to the membrane containing the reaction center.
The electron microscopy resolution is generally not sufficient to
distinguish the top and the bottom of the FMO trimer nor the
distance between FMO-RC. Thus, in principle either pairs of BChls 1
and 6 or BChls 3 and 4 are the pigments that connect the FMO complex
to RC. However, it is widely believed, due to efficient energy
funneling toward RC, that BChl 1 and 6 are the linkers to antenna
baseplate, and 3 and 4 should serve as target regions within the
neghibourhood of RC complex \cite{Renger06}. This hypothesis has
been recently verified experimentally \cite{Wen09}.  Up to this
point, we have considered BChl 3 to be in the close proximity of RC
by a trapping time-scale of about $1$ $ps$.  However, in this
section we consider both of these parameters to be free, in order to
explore the optimality and robustness of the ETE landscape as we
vary the time-scale and geometrical constraints set by the RC
trapping mechanism.

In Fig. \ref{Figefftrapcorrect}, we study the behavior of energy
transfer efficiency landscape in various trapping time-scales and
reorganization energies. It is evident that as the trapping rate
becomes very slow comparable to $100$ $ps$ or slower, the ETE drops
significantly independent of the values of $\lambda$. This can be
understood intuitively as follows: the excitation has to wait on
average so long for successful trapping to take place such that
there will be an increasing chance of electron-hole recombination as
we are approaching time-scales comparable to exciton life time. If
the trapping mechanism occurs within a time-scale of $1$ $ps$, the
ETE reaches to its expected maximum value of about 99\%. Generally,
one might expect that with increasingly faster trapping mechanisms
the likelihood of dissipation to environment vanishes and energy
transport approaches to the ideal case of having perfect efficiency.
However, when the trapping rate becomes very fast on the order of
$10^{-2}ps$ or faster, the ETE also drops significantly, a result
that might appear counter-intuitive.   In fact, overly rapid
trapping leads to low efficiency via the quantum Zeno effect, as the
rapid trapping effectively freezes the exciton dynamics and prevents
it from entering the reaction center. As a result, the finite
exciton life-time eventually leads to complete dissipation of
excitation to the environment in extreme limit of fast trapping of
about $1fs$.

\begin{figure}[tp]
\includegraphics[width=9cm,height=6cm]{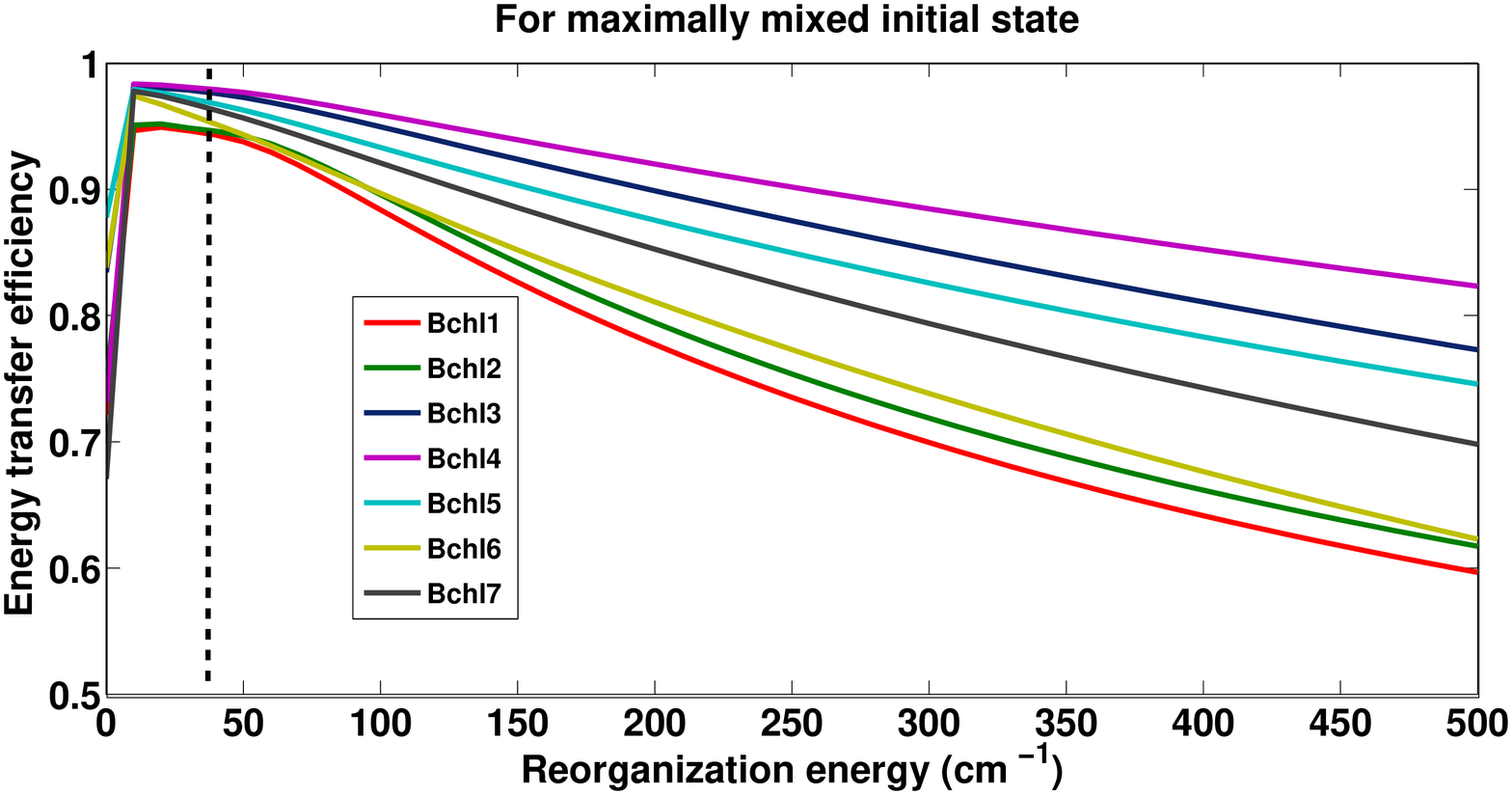}
\caption{The ETE as a function of reorganization energy for the
initially maximally mixed state. In each plot one of the 7 BChls is
considered to be connected to the reaction center. BChls 3 and 4
acting as the exciton transfer bridge yield the highest efficiency
for almost all values of the reorganization energy. This confirms
the experimental evidence that the FMO spatial orientation is such
that the BChls 3 and 4 are located near the RC.} \label{Figsites}
\end{figure}

The optimality and robustness of ETE versus both decoherence and
trapping rates using gradient and Hessian norms are presented in
Fig. \ref{Figopt-sus-lamtrap} (left panel). It can be observed that
at $\lambda=35 cm^{-1}$ for the FMO, the ETE is optimal with a
trapping rate of about 0.5ps. If the environmental interactions were
stronger, a comparably faster trapping mechanism would be required
to preserve such high level of efficiency.  However, for small and
intermediate system-bath interaction strength, where
environment-assisted transport occurs, slower trapping rates become
optimal, that is $\lambda$ and $r_{trap}$ are not competing
processes anymore. This implies that a multi-parameter convergence
of time-scales of the relevant physical processes might be required
for light-harvesting complexes to operate optimally \cite{mohseni13}. From Fig.
\ref{Figefftrapcorrect}, it can be easily seen that ETE is very
robust to variation of trapping rate at about $1$ $ps$ time scale.
In Fig. \ref{Figopt-sus-lamtrap}, we also illustrate the robustness
with respect to both trapping and reorganization energy (right
panel). For rather large values of $\lambda$, there are certain
regions that are highly robust to both parameters, but they are in
fact suboptimal, as can be seen from noting their values in the left
panel.

\begin{figure}[tp]
\includegraphics[width=9cm,height=6cm]{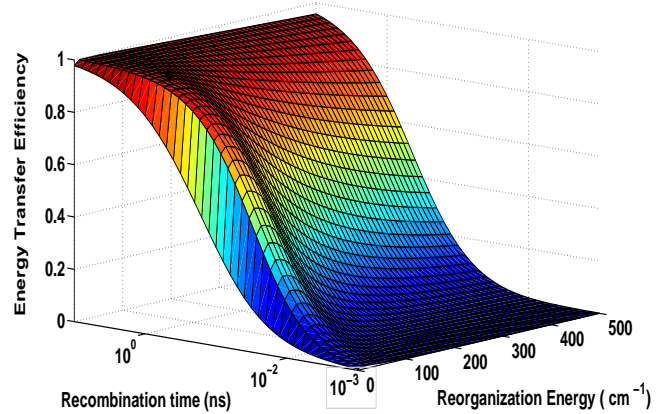}
\caption{ETE versus dissipation (loss) time-scale $r_{loss}^{-1}$
and reorganization energy. The maximum optimality and robustness for
FMO is observed around the estimated value $r_{loss}^{-1}=1ns$
implying the significance of the time-scale separation between
dissipation and trapping rates. We note that the ENAQT effect is
ubiquitous at all rates of electron-hole recombination process.}
\label{efflamrecomb}
\end{figure}

To explore the dependence of ENAQT effect on the location of
reaction center, we consider the efficiency of other scenarios that
the reaction center can be in the proximity of any other BChl sites.
Figure \ref{Figsites} shows ETE versus reorganization energy, with a fixed $\gamma=50 cm^{-1}$, for all possible
trapping sites. To be unbiased with the respect to the initial
state, we assume a maximally mixed initial state. It can be seen
that the optimal curves belong to BChls 3 and 4 as expected, since
they contribute highly to the lowest energy delocalized excitonic
states. It should be noted that optimal environment-assisted quantum
transport, and the two extreme regimes of quantum localization can
be observed for all of these plots independent of the actual
location of trapping.  In other words, the behavior of the energy
transport efficiency landscape and its dependence on a single
governing parameter are not properties of a particular choice of
trapping site in the FMO structure.

For completeness, we also investigate the ETE landscape as a
function of dissipation (loss) rate and reorganization energy in
Fig. \ref{efflamrecomb}. In our simulation of the FMO dynamics we
have used the estimated value of $r_{loss}^{-1}=1ns$. In Fig.
\ref{efflamrecomb}, however,  we treat loss rate as a free parameter
and we observe that for any stronger dissipation process, ETE would
have a suboptimal and less stable behavior. Thus, even if all the
other important parameters are within the optimal regime, a large
time-scale separation between dissipation and trapping rate is still
required to guarantee the highest performance for light-harvesting
complexes. Fig. \ref{efflamrecomb} also demonstrates that the
existence of ENAQT is independent of a particular choice of
dissipation rate.

%\begin{widetext}
\begin{figure}[tp]
\includegraphics[width=9cm,height=6cm]{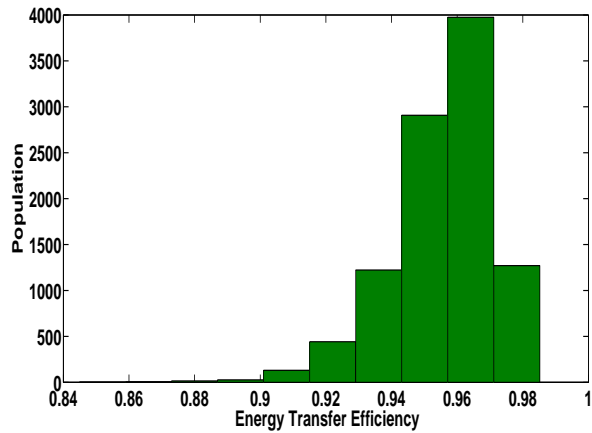}
\caption{Robustness of FMO transport efficiency with respect to
small variations of BChls locations, site energies and dipole
orientations for 10000 samples: the Hamiltonian parameters are
perturbed with site energies disorders of $\pm 10 cm^{-1}$,
dipole-moment uncertainties of $\pm 5^\circ$, and BChls spatial
displacement of $\pm 2.5 \AA$. The statistical distribution of
$10^4$ random configurations shows a significant degree of
robustness such that $99\%$ of samples still preserve an efficiency
of above $0.9$.} \label{Figconservative}
\end{figure}
%\end{widetext}

\section{robustness and optimality with respect to parameters of FMO
Hamiltonian}

So far, we have demonstrated that for the estimated Hamiltonian
elements of the FMO complex, the environmental parameters and
trapping rates are within the right set of values leading to an
optimal noise-assisted energy transfer efficiency. Moreover, the
performance of FMO is robust with respect to variations in such
decohering and lossy processes and to uncertainties in initial
conditions.  However, it is not fully clear if the FMO internal
Hamiltonian parameters have evolved to function optimally and fault
tolerantly, despite disorders and thermal fluctuations.  This issue
has been examined for LHCII in Ref. \cite{Sener02} using
semi-classical Pauli master equations to simulate the exciton
dynamics. Here, we would like to use TC2 to explore how rare is the
FMO geometry in terms of its efficiency, whether the specific
spatial and dipole moment arrangements of BChls are essential for
such highly efficient functioning of this pigment-protein complex,
and how robust these parameters are with respect to small and large
perturbations in chromophoric distances, dipole moment orientations,
and site energy fluctuations. Specifically, we explore if the FMO
closely packed structure plays any functional role, and illustrate a
potentially important convergence of the relevant dynamical
time-scales in the FMO energy transport. In the following section,
we investigate the underlying physical principle of quantum
transport in more generic multichromophoric structures beyond the
FMO geometry.

The Frenkel exciton Hamiltonian for a multichromophoric system is expressed as:
\begin{eqnarray}
H_{S} &=&\sum_{j,k}\epsilon _{j}|j\rangle \langle
j|+J_{j,k}|j\rangle
\langle k|,
\end{eqnarray}
in which $J_{jk}$ are Coulomb couplings of the
transition densities of the chromophores,
\begin{eqnarray}
J_{jk}\sim \frac{1}{R_{jk}^{3}}(\mathbf{\mu }_{j}\cdot \mathbf{\mu }%
_{k}-\frac{3}{R_{jk}^{2}}(\mathbf{\mu }_{j}\cdot \mathbf{R}_{jk})(\mathbf{%
\mu }_{k}\cdot \mathbf{R}_{jk})),
\end{eqnarray}
where $\mathbf{R}_{jk}$ denotes the distance between site $j$ and
$k$, and $\mathbf{\mu }_{j}$ is the transition dipole moment of
chromophore $j$ \cite{Damjanovi97}. We first study the robustness of
free Hamiltonian parameters within the proximity of the estimated
values for FMO as given in the appendix A.

\begin{figure}[tp]
\includegraphics[width=9cm,height=6cm]{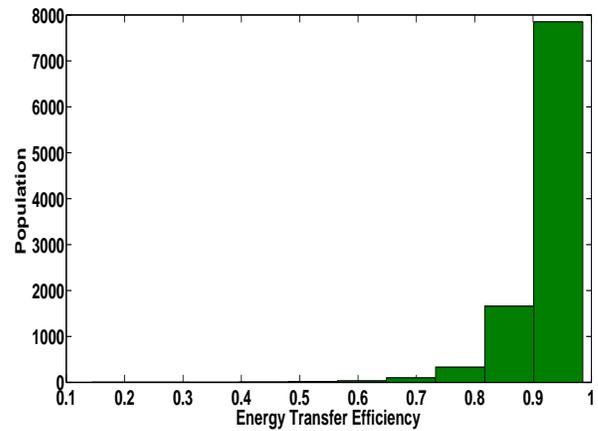}
\caption{Robustness of the FMO complex transport efficiency with
respect to large variations in BChls site energies and dipole moment
orientations for 10000 sample configurations: While the location of
BChls are still slightly perturbed, similar to Fig.
\ref{Figconservative} of about $\pm 2.5 \AA$, the dipole moments can
take any arbitrary direction and site energy takes any value between
zero and $500 cm^{-1}$. This Histogram reveals that the relative
distance of BChls is playing a crucial role in performance of these
random light-harvesting complexes since $79\%$ of them still hold
ETE larger than $90 \%$. } \label{Figrandom_angle-site}
\end{figure}

\subsection{Robustness of FMO Hamiltonian}

Figures \ref {Figconservative} and \ref{Figrandom_angle-site}
demonstrate the robustness of the FMO structure to variations in its
internal parameters. Figure \ref{Figconservative} illustrates that FMO
efficiency does not drop drastically with respect to variations
in the dipoles orientations, site energies, and Bchls
distances close to the neighborhood of the estimated values.
Specifically, from $10000$ random samples of FMO with spatial
uncertainty around each Bchl location of about $\pm 2.5$ $\AA$, dipole
moments orientations variations of $\pm 5^\circ$, and site energy
static disorder $\pm 10$ $cm^{-1}$, $97\%$ of configurations have
efficiency of $95\%$ or higher. This demonstrates a significant
degree of robustness with small perturbations. In order to separate
the influence of spatial coordinates from angular dipole
orientations and disorders, we allow the latter two parameters to
take arbitrary values from a large range while keeping Bchl locations uncertainties
to be limited by $\pm 2.5$ $\AA$. We observe in Fig.
\ref{Figrandom_angle-site} that ETE remains relatively robust with
$79\%$ of random $10000$ configurations have still efficiency of
$90\%$ or higher. This is rather counterintuitive considering huge
freedom that we have accommodated in the dipole moment arrangements
and site energies. These results indicate that spatial degrees of
freedom is a dominating geometrical ingredient of the FMO structure
and might play a key physical role in its performance. Similar robustness to the FMO system to variation in its structure had been independently reported in Ref. \cite{Jesenko}.

\subsection{ENAQT in presence of FMO-size variations}

\begin{figure}[tp]
\includegraphics[width=9cm,height=6cm]{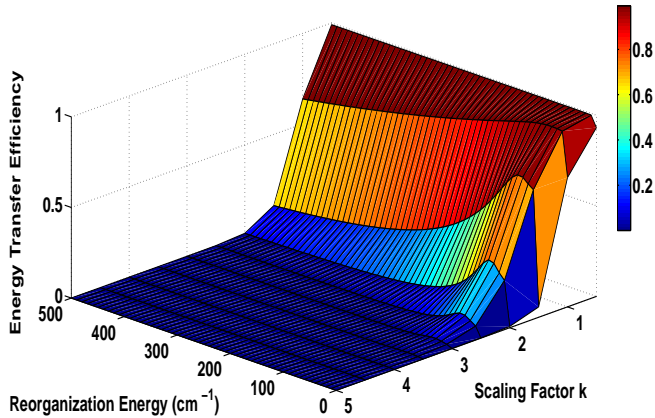}
\caption{Dependency of ETE on compactness level of the FMO complex.
The FMO chromophoric spatial structure is scaled with a factor
between $0.5$ and $5$. The ETE manifold is plotted as a function of
the scaling factor and reorganization energy. For all levels of
compactness, $\lambda=30-40 cm^{-1}$ yields the highest ETE. A more
compact complex shows a higher degree of robustness with the respect
to variation in reorganization energy. The ENAQT behavior can be observed at all levels of compactness, and almost all at the same reorganization energy value.} \label{FigcompactnessFMO}
\end{figure}

Our results presented in Figs. \ref{Figconservative} and
\ref{Figrandom_angle-site} clearly indicate that the relative Bchl
locations play a major role in the overall performance of the FMO
complex. Thus a potentially significant parameter of relevance is the
compactness of a given pigment-protein complex. We further explore
this feature by introducing a single \emph{compactness} parameter by
rescaling the Bchl distances by a factor k. We plot ETE as a
function of compactness level k varying by an order of magnitude
from $0.5$ to $5$. To explore any potential interplay of
environmental interactions with this particular internal degree of
freedom, we also simulate this size dependent
supersssion/enhancement of ETE in various reorganization energy,
$\lambda$, in Fig. \ref{FigcompactnessFMO}. It can be seen that
although transport efficiency drops significantly by expanding the FMO
structure, but the ENAQT
phenomenon remains scale invariant for the FMO-like structure.

\section{Conclusion}

In this paper, we report on robustness properties of ENAQT phenomenon therefore a complement to studies on the role of ENAQT in optimal transport. 
We considered excitonic energy transfer in FMO complex and presented
a comprehensive landscape study of ETE as a function of energy and
geometrical parameters representing system and environment degrees of freedom.  
We found that ENAQT can assist optimal quantum transport to be robust with respect to variations in system-environment parameters.  
Furthermore, we found that ENAQT has a universal behavior meaning that it does not disappear when some of parameters are out of optimal regime. 
The robustness of ENAQT is crucial when designing quantum transport systems enhanced by bath engineering. Here our observations is based on
numerical simulations of one quantum transport system, the FMO complex. It would be interesting and important to perform similar study for other
natural or artificial system.

\begin{acknowledgments}
We also thank J. H. Choi and D. Hayes for
helping us with extracting the FMO data. We acknowledge funding from
DARPA under the QuBE program (MM, AS, SL, HR), ENI (MM,SL), NSERC
(MM) and NSF (SL, AS,HR), and ISI, NEC, Lockheed Martin, Intel (SL).

\end{acknowledgments}

\appendix

\section{FMO Structure Information}

In this work we use the FMO Hamiltonian given in Ref.~\cite{Cho05}:
\[
H =
 {\begin{pmatrix}
 280 & -106 & 8 & -5 & 6 & -8 & -4  \\
 -106 & 420 & 28 & 6 & 2 & 13 & 1  \\
 8 & 28 & 0 & -62 & -1 & -9 & 17\\
 -5 & 6 & -62 & 175 & -70 & -19 & -57\\
 6 & 2 & -1 & -70 & 320 & 40 & -2\\
 -8 & 13 & -9 & -19 & 40 & 360 & 32\\
 -4 & 1 & 17 & -57 & -2 & 32 & 260
 \end{pmatrix} }
\]

Note that the coupling between BChls 5 and 6 given here is disputed to be compatible with the dipole-dipole approximation \cite{Renger06}.

Table I gives the estimated values of dipole moment orientations and positions of
Bchls, extracted from
the pdb file of the FMO complex \cite{Chopriv}.

\begin{table}[ht]
\caption{Spatial location of Bchls and their dipole moment
orientation.} \centering
\begin{tabular}{c c c c c c}
\hline\hline Bchl & x ($\AA$) & y ($\AA$) & z ($\AA$) & $\theta$ &
$\phi$ \\ [0.5ex] \hline
 1 & 28.032 & 163.534 & 94.400 &  0.3816 & -0.6423+$\pi$ \\ \hline
    2 & 17.140 & 168.057& 100.162& 0.067& 0.5209+$\pi$ \\ \hline
     3 &5.409& 180.553&  97.621& 0.1399& 1.3616+$\pi$ \\ \hline
    4&  9.062& 187.635 & 89.474 &  0.257 &-0.6098+$\pi$\\ \hline
    5&  21.823& 185.260 & 84.721& -0.1606& 0.6899+$\pi$  \\ \hline
    6&  23.815 &173.888&  82.810&  -0.4214& -1.4686+$\pi$ \\ \hline
    7&  12.735& 174.887&  89.044& 0.578& -1.0076+$\pi$ \\ [1ex]
\hline
\end{tabular}
\label{fmo_data}
\end{table}
The FMO inter-chlorophyll coupling is dipole-dipole interaction
\begin{eqnarray}
J_{jk}=\frac{C}{R_{jk}^{3}}(\mathbf{\mu }_{j}\cdot \mathbf{\mu }%
_{k}-\frac{3}{R_{jk}^{2}}(\mathbf{\mu }_{j}\cdot \mathbf{R}_{jk})(\mathbf{%
\mu }_{k}\cdot \mathbf{R}_{jk})),
\end{eqnarray}
for which we consider the constant $C|\mu|^2=134000$ $cm^{-1}\AA^3$ \cite{Renger06}.

\end{document}